\documentclass{jfm}

\usepackage{graphicx}
\usepackage{newtxtext}
\usepackage{newtxmath}
\usepackage{natbib}
\usepackage{hyperref}
\hypersetup{
    colorlinks = true,
    urlcolor   = blue,
    citecolor  = blue,
}

\newcommand{\RomanNumeralCaps}[1]

\title{Breaking bubbles across multiple timescales in turbulence}

\author{Yinghe Qi\aff{1}, Xu Xu\aff{1}, Shiyong Tan\aff{1}, Shijie Zhong\aff{1}, Qianwen Wu\aff{1}, 
 \and Rui Ni\aff{1}\corresp{\email{rui.ni@jhu.edu}}}

\affiliation{\aff{1}Department of Mechanical Engineering, Johns Hopkins University,
Baltimore, MD 21218, USA}

\begin{document}
\maketitle

\begin{abstract}
The familiar process of bubbles generated via breaking waves in the ocean is foundational to many natural and industrial applications. In this process, large pockets of entrained gas are successively fragmented by the ambient turbulence into smaller and smaller bubbles. The key question is how long it takes for the bubbles to reach terminal sizes for a given system. Despite decades of effort, the reported breakup time from multiple experiments differs significantly. Here, to reconcile those results, rather than focusing on one scale, we measure multiple timescales associated with the process through a unique experiment that resolves bubbles’ local deformation and curvature. The results emphasize that the scale separation among various timescales is controlled by the Weber number, similar to how the Reynolds number determines the scale separation in single-phase turbulence, but shows a distinct transition at a critical Weber number.
\end{abstract}

\begin{keywords}
Authors should not enter keywords on the manuscript, as these must be chosen by the author during the online submission process and will then be added during the typesetting process (see \href{https://www.cambridge.org/core/journals/journal-of-fluid-mechanics/information/list-of-keywords}{Keyword PDF} for the full list).  Other classifications will be added at the same time.
\end{keywords}

{\bf MSC Codes }  {\it(Optional)} Please enter your MSC Codes here

\section{Introduction}

The mesmerizing power of breaking ocean waves has long captivated the human imagination, but beneath the surface lies a complex two-phase flow problem as large pockets of gas being broken into a cloud of small bubbles by the ambient turbulence. This captivating process unfolds, substantially amplifying interfacial area, which, in turn, serves as a vital catalyst for enhanced gas transfer into the ocean \citep{deane2002scale,deike2016,lohse2018bubble,deike2022mass,gao2021bubble}.  In this process, the energy cost scales with the product of the power and the time it takes for bubbles to reach the terminal size through successive breakups. The power of the process can be estimated using the turbulence energy dissipation rate, which is directly related to the targeted bubble size based on the critical Weber number as proposed in the classic Kolmogorov-Hinze framework \citep{kolmogorov1949breakage,hinze1955fundamentals}. The timescale, however, was not discussed until it was first brought up in the seminal work by \cite{levich1962physicochemical}.

In Levich's work, the timescale associated with bubble breakup in turbulence can be characterized by examining the balance among the viscous force, pressure, and surface tension. Following this work, the breakup timescale is commonly considered to be dominated by external flows. If it is further assumed that the bubble size is the relevant length scale as hypothesized in the Kolmogorov-Hinze framework, the only timescale associated with the breakup process is the turn-over time of the bubble sized eddy. However, this hypothesis has been challenged recently \citep{qi2022fragmentation,vela2021deformation}, and it is found that sub-bubble-scale eddies may also contribute to the breakup process, resulting in a shorter timescale. In addition to the eddy timescale, more timescales relevant to the breakup process have been proposed \citep{ni2024deformation}, including the bubble natural oscillation timescale \citep{lamb1879hydrodynamics,risso1998oscillations}, the capillary timescale \citep{villermaux2020fragmentation,riviere2022capillary,ruth2022experimental}, the bubble life time \citep{qi2020towards,martinez1999breakup1,liao2009literature,vela2022memoryless}, the large-scale shear timescale \citep{zhong2023breakup}, and the convergent timescale \citep{qi2020towards,gaylo2023fundamental}. However, the relationship among those timescales remains unclear, particularly how this relationship is connected to the Weber number $We=\rho\langle\epsilon\rangle^{2/3} D^{5/3}/\sigma$, where $\rho$, $\langle\epsilon\rangle$, $D$, and $\sigma$ are the density, turbulence dissipation rate, bubble diameter, and surface tension coefficient, respectively.

So far, most studies of bubble breakup primarily focus on large-scale quantities such as the size \citep{deane2002scale,rodriguez2003novel,hesketh1987,vejrazka2018,yi2022physical,yi2023recent}, the aspect ratio \citep{masuk2021simultaneous,masuk2021towards,kang1989numerical,ravelet2011dynamics,lu2008effect,stone1986experimental}, the total surface area \citep{dodd2016interaction,legendre2012deformation}, and the low-order modes of spherical harmonics of the bubble \citep{perrard2021bubble,magnaudet2003drag} without considering the small-scale local interfacial deformation which could carry important clues on how bubbles react to the collision with sub-bubble-scale eddies. Therefore, in this work, we report the experimental measurements of the bubble local deformation by leveraging the full 3D reconstruction of bubble geometry. With the information of the deformation, multiple timescales associated with the breakup process are identified and their connection to the Weber number is discussed.

\begin{figure}
    \centering
    \includegraphics[width=\linewidth]{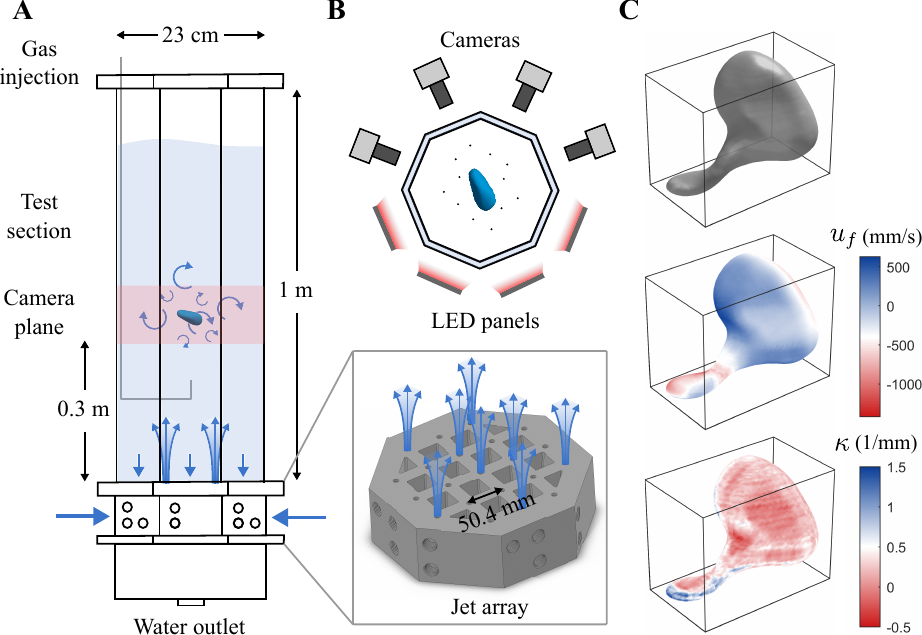}
    \caption{(a) A schematic of the vertical water tank as described in Appendix \ref{appA}. The blue arrows represent the direction of the flow. Inset: a 3D model of the jet array system for turbulence generation. (b) Top view of the configuration of cameras and LED panels around the octagonal test section. The black dots in the test section represent tracer particles around the bubble. (c) From top to bottom panel: the 3D reconstruction of a breaking bubble using the visual hull method; the distribution of the interfacial velocity $u_f$ for the same bubble; the distribution of the interfacial curvature $\kappa$ for the same bubble.}
    \label{fig1}
\end{figure}

\section{Results}\label{sec_result}

\begin{figure}
    \centering
    \includegraphics[width=\linewidth]{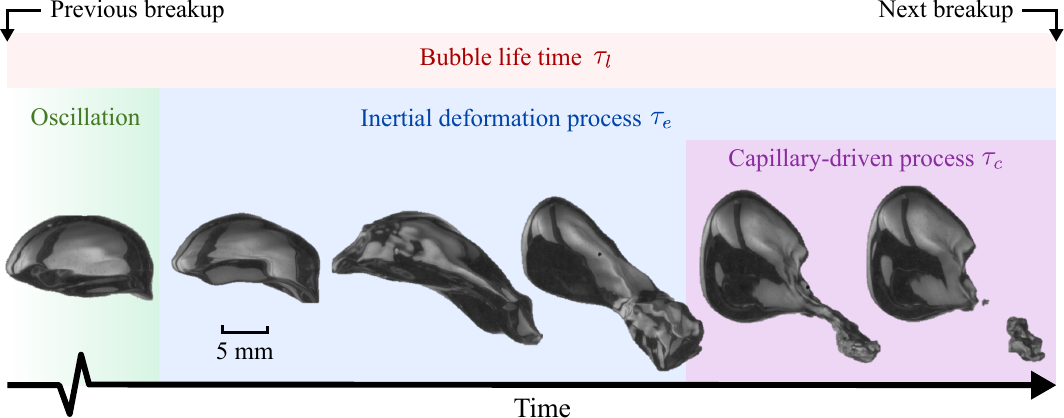}
    \caption{A sequence of snapshots of a breaking bubble with $D=10$ mm. The green, blue, purple, and red shaded area represents the oscillation process, the inertial deformation process, the capillary-driven process, and the bubble life time, respectively.}
    \label{fig2}
\end{figure}

Fig. \ref{fig1}(a) shows the schematic of the experimental apparatus in which homogeneous and isotropic turbulence (HIT) was generated through a jet array (inset). Bubbles were injected to the view volume via a needle connected to a gas line (gray line). A typical breakup process of a bubble is illustrated in Fig. \ref{fig2}. Four cameras from different angles with back lighting around the view volume were used (Fig. \ref{fig1}(b)) to obtain the 3D Lagrangian trajectories of tracer particles \citep{tan2020} and also to reconstruct the 3D geometry of bubbles \citep{masuk2019robust}. An example of the reconstructed bubble is shown in the top panel of Fig. \ref{fig1}(c). To better quantify the local deformation, the bubble interfacial velocity $u_f$ and mean curvature $\kappa$ are also calculated (as illustrated in Fig. \ref{fig1}(c)). More details regarding the experimental setup and the bubble interfacial velocity and curvature can be found in Appendix \ref{appA} and \ref{appB}. To further expand the parameter space, two supplemental datasets conducted in similar conditions \citep{masuk2021simultaneous,qi2022fragmentation} are also included.  Key parameters of all the datasets are summarized in Table \ref{tab_experiments}. In total, 385 breakup events (131 for Exp A, 183 for Exp B, and 40 for Exp C) are used for statistics of this work. Note that the breakup events are rare in the experiments with the mean Weber number close to $\mathcal{O}(1)$.

\begin{table}
  \begin{center}
  \begin{tabular}{lccccccc}  
       Experiment  & $\langle\epsilon\rangle$ (m$^2/$s$^3$) & $\eta$ ($\mu$m) & $\tau_\eta$ (ms) & $L$ (mm) & $D$ (mm) & $We$ & $Re_\lambda$ \\[3pt]
       Exp A (current experiment)   & 0.10 & 56 & 3.2 & 60 & 7--13 & 1.5--4.3 & 400\\
       Exp B \citep{masuk2021simultaneous}   & 0.16 & 50 & 2.5 & 60 & 3--9 & 0.5--1.6 & 435\\ 
       Exp C \citep{qi2022fragmentation}  & 0.2 & 47 & 2.2 & 15 & 2--4 & 0.3--1.0 & 180\\ 
  \end{tabular}
  \caption{Summary of parameters and dimensionless numbers of the experimental datasets included in the current work (Exp B and Exp C are the supplementary datasets). Here, $\eta$, $\tau_\eta$, $L$ and $Re_\lambda$ are the Kolmogorov length scale, Kolmogorov timescale, integral length scale and Taylor scale Reynolds number, respectively.}
  \label{tab_experiments}
  \end{center}
\end{table}

\begin{figure}
    \centering
    \includegraphics[width=\linewidth]{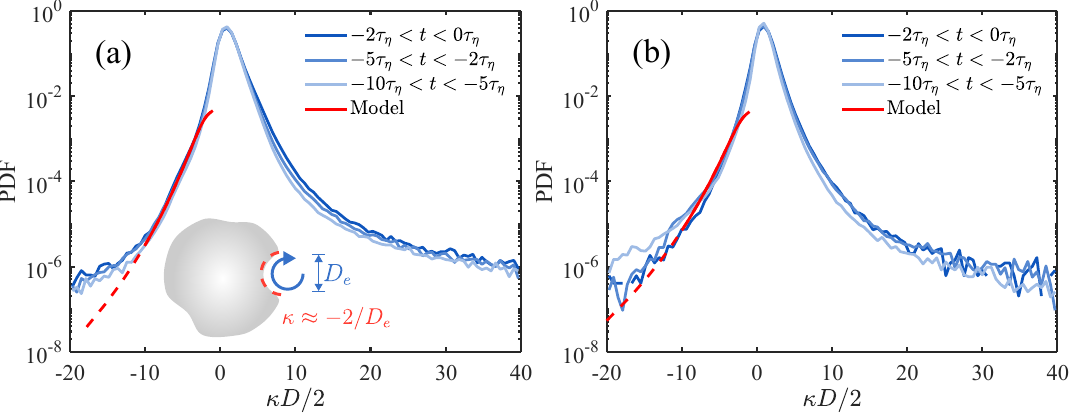}
    \caption{(a) PDFs of the normalized interfacial curvature $\kappa D/2$ for all breaking bubbles in Exp A. Blue colors represent different time before the breakup. The red line represents the prediction based on the model (Eq. \ref{eqn_indent_curv}). The solid part of the red line marks the range of $\kappa$ where the model is valid. (b) The same figure for Exp B.}
    \label{fig_curv_pdf}
\end{figure}

\subsection{Interfacial curvature}

Fig. \ref{fig_curv_pdf}(a) shows the probability density functions (PDFs) of the interfacial curvature $\kappa$ normalized by the bubble radius $D/2$ for bubbles that eventually break in Exp A. Here, $\kappa>0$ and $\kappa<0$ represent convex and concave interface, respectively. $t=0$ indicates the instant when the breakup occurs, and $t<0$ represents the moment before the bubble breakup. $\tau_\eta$ is the Kolmogorov timescale. The PDFs for the entire duration, from 10$\tau_\eta$ prior to the breakup to the breakup moment, follow a similar distribution, although a small difference over time can be observed. The peaks of all the PDFs are located at $\kappa D/2 \approx 0.9$, which is close to the undeformed spherical geometry with the mean curvature exactly at $\kappa D/2=1$. In addition, all the PDFs are positively skewed with a higher probability of finding interfaces with local positive curvatures, i.e. extruded outward from the gas phase to the ambient liquid. This positive skewness is potentially due to the incompressibility of the inner gas. As part of the bubble interface is compressed by surrounding turbulence (which will be discussed in the following), the rest of the interface tends to extrude outward. This extrusion might not be uniform but significantly sharp along certain directions with minimum resistance, resulting in $\kappa\gg 0$.

Note that the negative $\kappa$ represents the local depression on the bubble interface, which is associated with the increase of local dynamic pressure. If this depression is considered as a result of the collision between the bubble interface and sub-bubble-scale eddies \citep{luo1996theoretical}, the left tail of the PDF can be understood and modeled by accounting for such interactions. 

Let us picture a simple scenario in which a bubble with a diameter $D$ encounters an energetic, sub-bubble-scale eddy with a size of $D_e<D$, as shown in the schematic of Fig. \ref{fig_curv_pdf}(a). To the leading order, the local curvature can be assumed to scale with $1/D_e$, and the other part of the interface remains unchanged given the short interaction time. In order for the eddy to depress the local interface, the inertia of the eddy $\rho u_e^2$ must be comparable with the surface tension induced by the local curvature, i.e., $\sigma/D_e$, where $u_e$ is the velocity of the eddy. This relation leads to $\rho u_e^2>C_1\sigma/D_e$, where $C_1$ is a fitting parameter that will be discussed in detail later. Rearranging this equation leads to the minimum eddy velocity required to deform the interface 
\begin{equation}
    u_{e,d}=\sqrt{C_1\sigma/(\rho D_e)}.
\end{equation}

The conditional PDF of instantaneous eddy velocity $u_e$ in turbulence for a given eddy size $D_e$ can be expressed as 
\begin{equation}\label{eqn_eddy_vel_pdf}
    P(u_e|D_e)=3\sqrt{2}\epsilon_e^{2/3} D_e^{-1/3}P(\epsilon_e)/2,
\end{equation}
where $\epsilon_e$ is the local energy dissipation rate at the eddy length scale \citep{qi2022fragmentation}. The distribution of $\epsilon_e$ can be approximated by a log-normal function $P(\epsilon_e)=1/(\epsilon_e\sqrt{2\pi\sigma_{\ln\epsilon}^2})\exp{\left[-\left(\ln{(\epsilon_e/\langle\epsilon\rangle)}+\sigma_{\ln\epsilon}^2/2\right)^2/{2\sigma_{\ln\epsilon}^2}\right]}$, given by the multi-fractal model \citep{meneveau1991multifractal,kolmogorov1962}. Here $\sigma_{\ln\epsilon}^2=A+\mu\ln(L/D_e)$ is the variance; $A$ represents a large-scale variability, which is set at $A=0$ for convenience; $\mu\approx0.25$ is the intermittency exponent; and $L$ is the integral length scale of turbulence.

Given the distribution of the instantaneous eddy velocity (Eq. \ref{eqn_eddy_vel_pdf}), the PDF of the size of the eddies which are sufficiently strong to depress the bubble interface can be expressed by 
\begin{equation}
    P(D_e)\sim D_e^2\omega_c \int_{u_{e,d}}^\infty P(u_e|D_e) du_e.
\end{equation}
This equation accounts for the interfacial area depressed by the bubble, which scales with  $\sim D_e^2$ (as illustrated in Fig. \ref{fig_curv_pdf}(a)), and the frequency of the collision between the bubble and an eddy of size $D_e$ \citep{qi2022fragmentation,luo1996theoretical}, which can be estimated using 
\begin{equation}\label{eq_collision_freq}
    \omega_c\sim \langle\epsilon\rangle^{1/3}D^2D_e^{-11/3}.
\end{equation}
Since the local curvature of the depression can be approximated by $\kappa\approx-2/D_e$, the PDF of the local negative curvature on the bubble interface can therefore be expressed as
\begin{equation}\label{eqn_indent_curv}
    P(\kappa)\sim\kappa^{-2}P(D_e).
\end{equation}

The predicted PDF for $\kappa<0$ based on Eq. \ref{eqn_indent_curv} is shown in Fig. \ref{fig_curv_pdf}(a) as the red solid curve, with only one fitting parameter $C_1=0.36$ to set the minimum eddy velocity $u_{e,d}$. It is evident that the model prediction agrees with the experimental results but only for a range of $\kappa$ because Eq. \ref{eqn_indent_curv} works for eddies only within the inertial range. As a result, the predicted PDF extends only up to $\kappa D/2\approx -10$, corresponding to the eddy size of $D_e\approx 1$ mm, which is close to the lower limit of the inertial range (see Appendix \ref{app_struct_func}). In addition, the model can only describe local deformation so it cannot be used to predict the distribution of small curvature that corresponds to large deformation of the bubble size, i.e. $\kappa D/2\approx -1$. Nevertheless, the overall agreement for the scales considered suggests that the bubble local deformation is likely driven by the collision with sub-bubble-scale eddies at least statistically. In Fig. \ref{fig_curv_pdf}(b), the analysis above is also conducted for Exp B for bubbles with a size of 3--6 mm, which is around half of the bubble sizes in Exp A. The PDFs follow similar distributions as those in Fig. \ref{fig_curv_pdf}(a), and a similar good agreement between the model and experiments for the range of scales considered can still be established by setting the fitting parameter $C_1=0.14$ in the model.

\begin{figure}
    \centering
    \includegraphics[width=\linewidth]{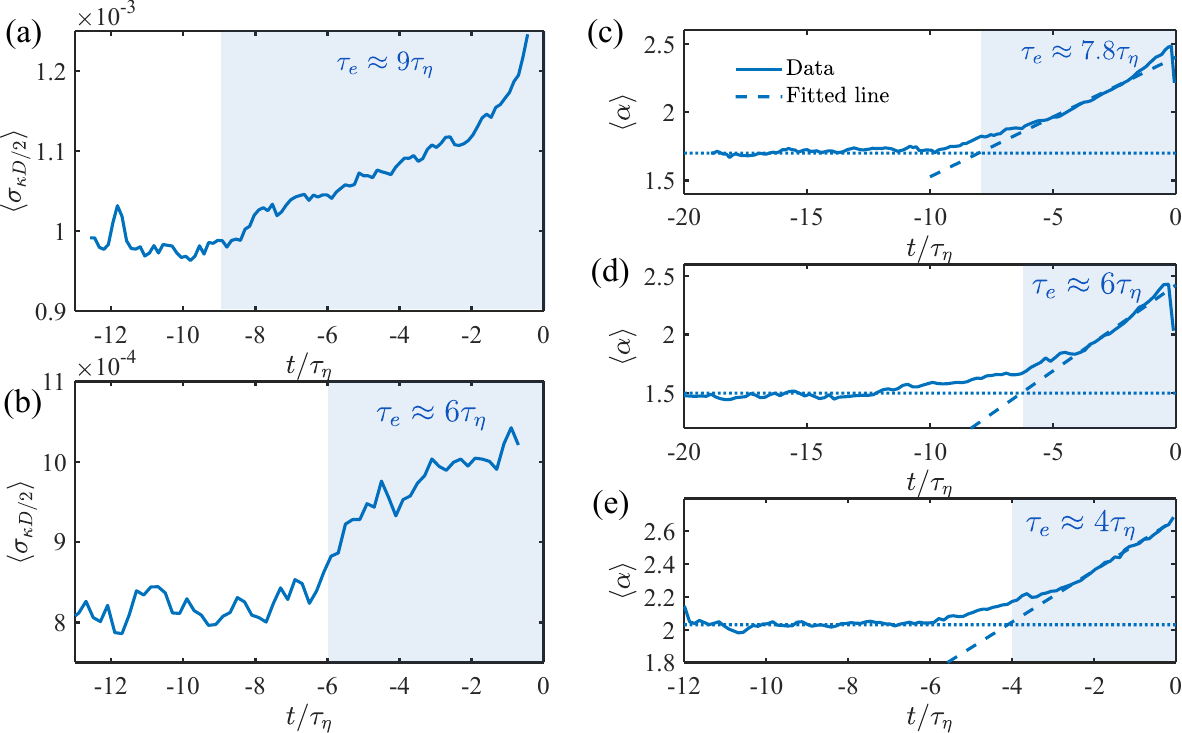}
    \caption{(a) The time evolution of the averaged standard deviation of the curvature $\langle\sigma_{\kappa D/2}\rangle$ for Exp A. The blue shaded area marks the timescale $\tau_e$. (b) The same figure for Exp B. (c, d, and e) The time evolution of the mean aspect ratio $\langle\alpha\rangle$ (solid lines) of breaking bubbles for Exp A, B, and C, respectively. The blue shaded area marks the timescale $\tau_e$.}
    \label{fig_curv_evo}
\end{figure}

Prior to the breakup, as shown in Fig. \ref{fig2} by the green shaded area, bubbles experience strong deformation, including frequent stretching and depression so the variation of curvature could be an indicator of the breakup dynamics. Fig. \ref{fig_curv_evo}(a) shows the time evolution of the averaged standard deviation of the curvature $\langle\sigma_{\kappa D/2}\rangle$ over all the breaking-bubble trajectories before the breakup occurs for Exp A. It is seen that $\langle\sigma_{\kappa D/2}\rangle$ remains roughly constant until around $t=-9\tau_\eta$ when $\langle\sigma_{\kappa D/2}\rangle$ begins to gradually increase, indicating a substantial variation of the interfacial curvature. The associated timescale $\tau_e\approx 9\tau_\eta$ as indicated by the blue shaded area is then considered as the characteristic timescale of such a turbulence-driven inertial deformation process \citep{ruth2022experimental} (blue shaded area in Fig. \ref{fig2}).

In addition to the curvature, another quantity which reflects the deformation of bubbles is the aspect ratio $\alpha$. Fig. \ref{fig_curv_evo}(c) shows the time evolution of the mean aspect ratio $\langle\alpha\rangle$ of all the breaking bubbles for the same dataset prior to the breakup moment. Here, the aspect ratio $\alpha=2r_\text{maj}/D$ is defined as the ratio between the semi-major axis $r_\text{maj}$ and the bubble spherical-equivalent radius $D/2$. In Fig. \ref{fig_curv_evo}(c), it is evident that $\langle\alpha\rangle$ far from the breakup moment remains almost constant with $\langle\alpha\rangle>1$, suggesting that bubbles experience strong temporal oscillation (as indicated by the green shaded area in Fig. \ref{fig2}). However, since those fluctuations do not follow the same frequency or phase, averaging the signals over many bubbles results in a nearly constant $\langle\alpha\rangle$. Close to the breakup moment, $\langle\alpha\rangle$ exhibits clear growth, implying that $\alpha$ for all bubbles is likely to increase during this time period. A method is applied to extract this timescale by calculating the intersection between two lines: the dotted line that captures the early plateau and the dashed line that is fitted over the range of $\langle\alpha\rangle$ from the half height to the peak, as illustrated in Fig. \ref{fig_curv_evo}(c). A timescale of around $7.8\tau_\eta$ can be then extracted as indicated by the blue shaded area. This timescale is similar to $\tau_e\approx 9\tau_\eta$ obtained from Fig. \ref{fig_curv_evo}(a) and is thus considered as the same timescale associated with the bubble inertial deformation driven by turbulence. 

The same analysis of $\langle\sigma_{\kappa D/2}\rangle$ and $\langle\kappa\rangle$ is also repeated for Exp B with 3--6 mm bubbles, as shown in Fig. \ref{fig_curv_evo}(b) and (c) respectively. Most of the previous discussion regarding Exp A remains the same here and a shorter inertial deformation timescale $\tau_e\approx 6\tau_\eta$ can be consistently obtained from the time evolution of both $\langle\sigma_{\kappa D/2}\rangle$ and $\langle\kappa\rangle$. For Exp C, since the bubble is too small so that the reliable reconstruction of the curvature is not possible, $\tau_e$ is only determined based on the time evolution of $\langle\alpha\rangle$ (Fig. \ref{fig_curv_evo}(e)).

\subsection{The inertial deformation timescale}

\begin{figure}
    \centering
    \includegraphics[width=0.8\linewidth]{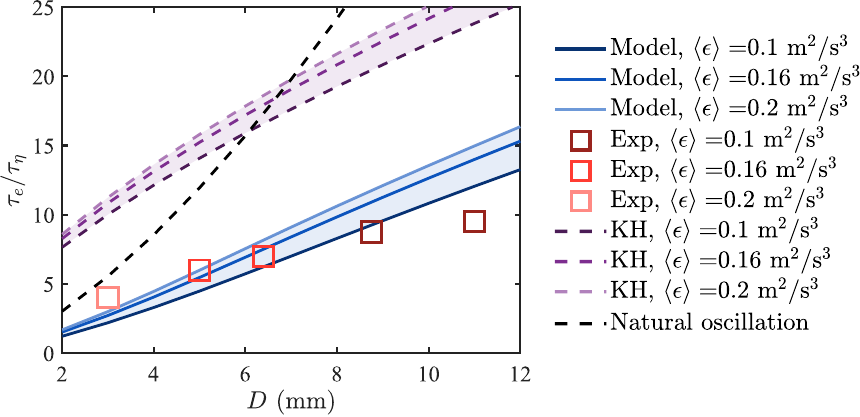}
    \caption{The inertial deformation timescale $\tau_e$ as a function of the bubble size. Squares represent the experimental data from all the datasets. The solid lines are the prediction based on the model (Eq. \ref{eqn_breakup_time}). The purple dashed lines are the turn-over time of the bubble-sized eddy. The black dashed line represents the second-mode natural oscillation timescale of the bubble. }
    \label{fig_inertial_timescale}
\end{figure}

Following a similar procedure as discussed above, the timescale of the inertial deformation, i.e. $\tau_e$, is extracted and shown in Fig. \ref{fig_inertial_timescale} for all the datasets. In the case when $\tau_e$ obtained based on the time evolution of $\langle\sigma_{\kappa D/2}\rangle$ and $\langle\kappa\rangle$ is slightly inconsistent (e.g., Fig. \ref{fig_curv_evo}(a) and (c)), the averaged $\tau_e$ is used.

Given the fact that $\tau_e$ is driven by surrounding turbulence, two simple approaches are available to model $\tau_e$. The first approach is the Kolmogorov-Hinze framework \citep{kolmogorov1949breakage,hinze1955fundamentals}, based on which the timescale of the inertial deformation process $\tau_e$ can be estimated using the turn-over timescale of the bubble-sized eddy, i.e., $\tau_D= D/(\sqrt{C_2}(\langle\epsilon\rangle D))^{1/3}$ which is shown in Fig. \ref{fig_inertial_timescale} as purple dashed lines for various $\langle\epsilon\rangle$. The other approach is to associate $\tau_e$ with the resonance oscillation of the bubble which could also lead to breakup \citep{risso1998oscillations}. The timescale of this resonance oscillation is given by $\tau_2=2\pi \sqrt{ \rho D^3/(96\sigma)}$ \citep{lamb1879hydrodynamics} which is shown in Fig. \ref{fig_inertial_timescale} as black dashed lines. In Fig. \ref{fig_inertial_timescale}, it is evident that both approaches overestimate $\tau_e$ compared to the experimental data.

The alternative way to model $\tau_e$ is by following a similar approach in the previous work by \cite{qi2022fragmentation}, in which the sub-bubble-scale eddy contribution to the breakup is incorporated and the breakup criterion is set by two relationships based on the inertia and timescale of the sub-bubble-scale eddy. Following this work, the minimum requirement of the eddy velocity $u_{e,b}$ (or equivalently the eddy kinetic energy) to break the bubble can be written as 
\begin{equation}
    u_{e,b}(D_e,D)=\max\left(\sqrt{\sigma/(\rho D_e)},\sqrt{96\sigma/(4\pi \rho D^3 D_e^{-2})}\right).
\end{equation}
Given $u_{e,b}$, considering the distribution of eddy velocity $P(u_e|D_e)$ of an eddy (Eq. \ref{eqn_eddy_vel_pdf}), the probability $P'_b$ for this eddy to break the bubble is expressed following the integration
\begin{equation}\label{eqn_break_prob}
    P'_b=\int_{u_{e,b}}^\infty P(u_e|D_e) du_e.
\end{equation}

In order to estimate the timescale, it is assumed that the time required for a bubble to be broken by an eddy with a velocity of $u_e$ is given by $D/(2u_e)$. The factor of 2 in the denominator comes from the observation that the formation of the neck typically involves retraction of the interface simultaneously from both sides. Considering the distribution of eddy velocity $P(u_e|D_e)$ (Eq. \ref{eqn_eddy_vel_pdf}), the breakup probability $P'_b$ (Eq. \ref{eqn_break_prob}) and the collision frequency $\omega_c$ (Eq. \ref{eq_collision_freq}), the expected inertial timescale, i.e., $\tau_e$, can be expressed by

\begin{equation}\label{eqn_breakup_time}
    \tau_e=\frac{\int^D_{10\eta} \int_{u_{e,b}}^\infty D/(2u_e) P(u_e|D_e) P_b' \omega_c du_e dD_e}{\int^D_{10\eta}P_b' \omega_c dD_e},
\end{equation}
where the contribution from all the sub-bubble-scale eddies in the inertial range, from around $\sim 10\eta$ ($\eta$ being the Kolmogorov length scale) to the bubble size, are incorporated. Eddies in the dissipation range are not considered here as they have negligible energy to break the bubble and Eq. \ref{eqn_eddy_vel_pdf} and \ref{eq_collision_freq} do not hold anymore. Note that, in this model, no free parameter is involved.

It is worth emphasizing the difference between Eq. \ref{eqn_breakup_time} and the bubble life time model reported in \cite{qi2022fragmentation}. The bubble life time model in \cite{qi2022fragmentation} considers the averaged time required for the bubble to encounter an eddy which eventually leads to a successful breakup, whereas Eq. \ref{eqn_breakup_time} is the weighted average of the time between a successful collision event and the final breakup.

The prediction based on Eq. \ref{eqn_breakup_time} for various $\langle\epsilon\rangle$ is shown as blue solid lines in Fig. \ref{fig_inertial_timescale}. Compared to the classical Kolmogorov-Hinze framework, it is evident that the predicted $\tau_e$ agrees better with the experimental data, further emphasizing the role of energetic, sub-bubble-scale eddies in the inertial deformation process. It is noted that the Kolmogorov-Hinze framework (purples lines) could also achieve reasonable agreement with the data by assuming the eddy size does not have to match the bubble size. Following this assumption, a coefficient $\alpha$ can be introduced to the length scale (i.e., $\tau_D= \alpha D/(\sqrt{C_2}(\langle\epsilon\rangle\alpha D))^{1/3}$) and its magnitude is found to be $\alpha<1$ based on Figure \ref{fig_inertial_timescale}. This result suggests that the inertial deformation timescale could potentially be dominated by the turn-over timescale of a sub-bubble-scale eddy, consistent with the underlying mechanism of Eq. \ref{eqn_breakup_time} although only one sub-bubble scale is considered here. It is also worth noting that Eq. \ref{eqn_breakup_time} slightly over-predicts $\tau_e$ at $D=11$ mm, and it is likely driven by the buoyancy effect. In particular, the Eötvös number $Eo=\rho g D^2/\sigma$ for such bubble size is around $Eo=16.8\gg1$. This effect likely destabilizes the bubbles and results in an accelerated breakup process, especially for large bubbles with relatively low turbulence intensity.


\subsection{Interfacial velocity}

\begin{figure}
    \centering
    \includegraphics[width=\linewidth]{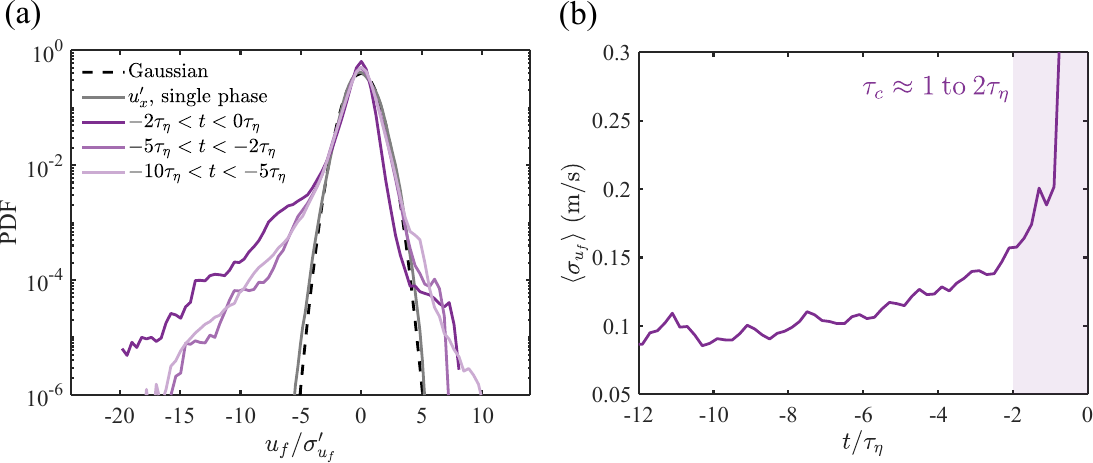}
    \caption{(a) PDFs of the interfacial velocity $u_f$ for 3--6 mm breaking bubbles in Exp B. Purple solid lines represent the different times before the breakup. The black dashed line marks the Gaussian distribution. The grey solid line shows the PDF of one component of the single-phase fluctuation velocity $u_x^\prime$ normalized by its standard deviation. (b) The time evolution of the averaged standard deviation of the interfacial velocity $\langle\sigma_{u_f}\rangle$. The purple shaded area marks the timescale $\tau_c$. }
    \label{fig_intf_vel_ashik}
\end{figure}

In addition to $\tau_e$, a different timescale can be extracted from the interfacial velocity statistics. Fig. \ref{fig_intf_vel_ashik}(a) shows the PDFs of the interfacial velocity $u_f$ for 3--6 mm breaking bubbles over different times in Exp B.  $u_f>0$ and $u_f<0$ represent outward and inward interfacial velocity respectively, from the perspective of bubbles. As a result, $u_f>0$ indicates interface moving from the gas phase towards the outer liquid phase, and vice versa. In this figure, $u_f$ is normalized by its standard deviation $\sigma_{u_f}'$. It is evident that the PDFs for $t<-2\tau_\eta$ collapse well and follow a similar distribution. However, the PDF at $-2\tau_\eta<t<0$ is slightly off with a higher left tail and a lower right tail. In addition, all the PDFs clearly deviate from a Gaussian distribution, which is shown in Fig. \ref{fig_intf_vel_ashik}(a) by the black dashed line. The much higher probability of finding extreme interfacial velocity implies that the interfacial velocity is more intermittent compared to the single-phase fluctuation velocity as shown in Fig. \ref{fig_intf_vel_ashik} by the grey solid line. Moreover, the PDFs of $u_f$ are all negatively skewed, indicating that the inward interfacial velocity is substantially stronger compared to the outward. This strong inward interfacial velocity could be linked to the distribution of the interfacial curvature, which shows clear positive skewness (see Fig. \ref{fig_curv_pdf}). As the majority of the interface is convex with $\kappa D/2\gg 1$, the resulting surface tension pointing inward from liquid to gas tends to significantly accelerate the contraction of the interface, leading to stronger negative interfacial velocity.

In addition to the PDFs, the time evolution of the standard deviation of the interfacial velocity, averaged over all the breaking-bubble trajectories $\langle\sigma_{u_f}\rangle$, is shown in Fig. \ref{fig_intf_vel_ashik}(b). $\langle\sigma_{u_f}\rangle$ remains almost constant at the beginning, indicating that the distribution of $u_f$ does not experience significant changes. At $t\approx-6\tau_\eta$, around one inertial deformation timescale earlier before the breakup ($\tau_e\approx6\tau_\eta$),  $\langle\sigma_{u_f}\rangle$ begins to grow. Such growth suggests the inertial deformation process of the bubble could affect the distribution of the interfacial velocity. Later on, at around $t\approx-2$ to $-1\tau_\eta$, another substantial increase of $\langle\sigma_{u_f}\rangle$ is observed. This sudden increase leads to a new timescale $\tau_c\approx 1$ to $2\tau_\eta$, as indicated by the purple shaded area in Fig. \ref{fig_intf_vel_ashik}(b). Note that this timescale $\tau_c$ is significantly shorter than the corresponding inertial deformation timescale ($\tau_e\approx 6\tau_\eta$), indicating a distinct physical process occurs right before the breakup.


The mechanism that leads to the sudden increase of $\langle\sigma_{u_f}\rangle$ is likely linked to the necking process before the neck finally pinches off as illustrated by the purple shaded area in Fig. \ref{fig2}. With sufficient amount of deformation, the neck shrinks rapidly right before the breakup, which is accompanied by a fast and local inward interfacial velocity. This interfacial velocity eventually leads to the sudden increase of $\langle\sigma_{u_f}\rangle$ in Fig. \ref{fig_intf_vel_ashik}(b) and the rise of the left tail of the PDF in Fig. \ref{fig_intf_vel_ashik}(a) during $-\tau_c<t<0$.

Based on the discussion above, the timescale $\tau_c$ of the final pinch-off can be modeled using the capillary timescale which has been discussed before \citep{riviere2022capillary,ruth2022experimental,villermaux2020fragmentation}. The model of the capillary timescale follows $\tau_c=\rho^{1/2} \sigma^{-1/2} \delta^{3/2}/(2\sqrt{3})$, where $\delta$ is the typical width of the bubble neck that eventually becomes unstable and leads to the rupture of the interface. $\delta$ can be estimated using the size of the smallest daughter bubble resulting from the breakup \citep{riviere2022capillary}. In our experiments, the neck width and the daughter bubble size vary from case to case. The mean neck width is estimated roughly on the order of $\delta\sim\mathcal{O}(1)$ mm for bubbles with $D\gtrsim 3$ mm. Using $\delta=1$ mm as the neck width leads to an estimation of $\tau_c\approx1.5\tau_\eta$ which is reasonably close to the range of $\tau_c$ measured from the experiment (Fig. \ref{fig_intf_vel_ashik}(b)). 

\subsection{Breakup timescales and separation}



Fig. \ref{fig2} illustrates two timescales associated with the breakup process, including the inertial deformation time $\tau_e$ and the capillary timescale $\tau_c$. In addition, the bubble life time $\tau_l$ which is defined as the time interval between two consecutive breakups is also shown, as indicated by the red shaded area. During this bubble life time, bubbles experience complicated deformation and oscillation due to the continuous bombardment by eddies of various sizes (green shaded area), until the event that the bubble encounters intense eddies, leading to strong inertial deformation followed by the neck thinning, which ultimately leads to the breakup. 

\begin{figure}
    \centering
    \includegraphics[width=0.75\linewidth]{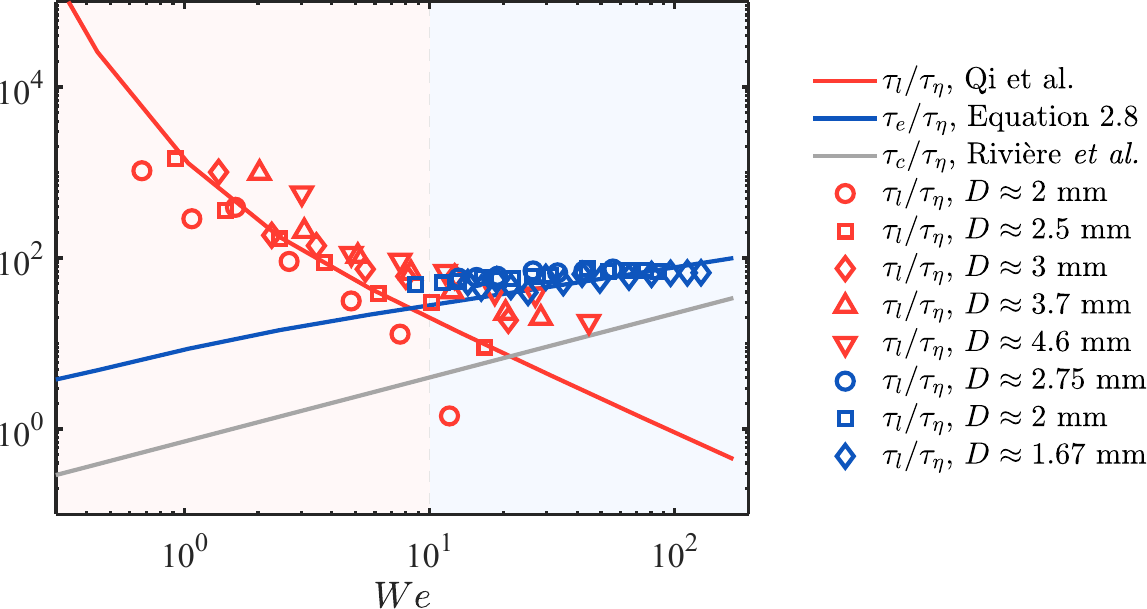}
    \caption{The comparison among different timescales associated with bubble breakup predicted by models, including the bubble life time $\tau_l$ (red solid line), the inertial deformation timescale $\tau_e$ (blue solid line), and the capillary timescale $\tau_c$ (gray solid line), as functions of $We$. The red and blue symbols represent the experimental data of $\tau_l$ by \cite{vejrazka2018} and \cite{martinez1999breakup1}, respectively. The red and blue shaded area marks $We<10$ and $We>10$.}
    \label{fig_time_compare}
\end{figure}

To further compare the bubble life time $\tau_l$ with the other timescales, the experimental data from previous works are included. Fig. \ref{fig_time_compare} shows the measured bubble life time $\tau_l$ by \cite{vejrazka2018} (red symbols) and \cite{martinez1999breakup1} (blue symbols) as a function of the Weber number. Here, $\tau_l$ is calculated following $\tau_l=1/g_b$, where $g_b$ is the bubble breakup frequency which can be measured experimentally \citep{haakansson2020validity}. Both datasets show distinct trends. Specifically, \cite{vejrazka2018} reported a decrease in bubble life time as $We$ increases, while \cite{martinez1999breakup1} suggested a slightly increasing $\tau_l/\tau_\eta$. This difference in the Weber number dependence marks the decades-long debate on whether the life time of bubbles increases or decreases as a function of $We$.

In addition to the experimental results, the solid red line in Fig. \ref{fig_time_compare} also shows the prediction by a bubble life time model \citep{qi2022fragmentation} for 3 mm bubbles at various $\langle\epsilon\rangle$. This model already considers the contribution from sub-bubble-scale eddies, i.e., bubble breakups are accelerated by the collisions with sub-bubble-scale, energetic eddies with the bubble life time dominated by the frequency of those extreme events. The model agrees well with the experimental data for $We\lesssim10$ however decays continuously and underestimates the life time for $We\gtrsim10$. Note that this model does not account for the fact that the life time $\tau_l$ cannot drop indefinitely as bubbles still need at least one inertial deformation timescale $\tau_e$ to complete the breakup process. Fig. \ref{fig_time_compare} shows the predicted $\tau_e$ by Eq. \ref{eqn_breakup_time} for 3 mm bubbles as the blue solid line. The result by Eq. \ref{eqn_breakup_time} is consistent with experimental data for $\tau_l$ at the large Weber number limit. Note that both experimental datasets can be explained by considering the maximum between the predicted bubble life time \citep{qi2022fragmentation} and the predicted inertial deformation timescale (Eq. \ref{eqn_breakup_time}), which cross over each other at $We\approx 10$. The agreement suggests that different trends observed in the experiments were the result of the transition of different timescales at play.

This result also highlights the role of the Weber number in determining the timescale separation for breakup in turbulent two-phase flows. For small $We$, since the majority of collisions between bubbles and surrounding eddies lead only to deformation but not breakup, the life time of the bubble is dominated by the long oscillation stage, and the scale separation between the life time and the final inertial deformation timescale is significant. Such scale separation depends only on $We$. For large $We$, the two timescales become one and no scale separation can be observed. This is in analogous to the role played by the Reynolds number in controlling the scale separation in single-phase turbulence where the separation between the Kolmogorov scale and integral scale diminishes as Re decreases. The capillary timescale $\tau_c$ \citep{riviere2022capillary} (gray solid line in Fig. \ref{fig_time_compare}), however, remains nearly a small constant regardless of $We$ as it is not affected by surrounding turbulence.

\section{Conclusion}\label{sec_conclusion}


The deformation and breakup of bubbles in turbulence involve the interaction between bubbles and eddies of various sizes. Which characteristic timescales control the breakup process is an inherently challenging question to answer and the framework by \cite{levich1962physicochemical} proposed several dominant timescales without specifying the relationships among them. In this work, to examine these timescales, we employed different statistics based on the shape reconstruction of breaking bubbles, including large-scale aspect ratio and small-scale local interfacial deformation through a unique experiment. The distribution of bubble local curvature emphasizes the importance of sub-bubble-scale eddies. Following this hypothesis, a small-eddy collision model to predict the interfacial curvature distribution is proposed and is found in good agreement with experimental results. A timescale associated with the inertial deformation of the bubble induced by the collision with sub-bubble-scale eddies and a capillary timescale associated with the neck process are then identified and modeled based on curvature and interfacial velocity statistics. In addition to these timescales, the bubble life time is also briefly discussed. It is found that for $We<10$, the bubble life time is dominated by the frequency of energetic sub-bubble-scale eddies that can break the bubble. However, for $We>10$, the majority of sub-bubble-scale eddies are sufficiently energetic and thus this timescale is limited primarily by the inertial deformation timescale. The results highlight how the Weber number controls the separation of timescales in the breakup dynamics, similar to the well-known role of the Reynolds number in determining the scale ratio in single-phase turbulence.

\backsection[Supplementary data]{\label{SupMat}Supplementary material and movies are available at \\https://doi.org/10.1017/jfm.2019...}

\backsection[Acknowledgements]{Acknowledgements may be included at the end of the paper, before the References section or any appendices. Several anonymous individuals are thanked for contributions to these instructions.}

\backsection[Funding]{Please provide details of the sources of financial support for all authors, including grant numbers. Where no specific funding has been provided for research, please provide the following statement: "This research received no specific grant from any funding agency, commercial or not-for-profit sectors." }

\backsection[Declaration of interests]{The authors report no conflict of interest.}

\backsection[Data availability statement]{The data that support the findings of this study are openly available in [repository name] at http://doi.org/[doi], reference number [reference number]. See JFM's \href{https://www.cambridge.org/core/journals/journal-of-fluid-mechanics/information/journal-policies/research-transparency}{research transparency policy} for more information}

\backsection[Author ORCIDs]{Authors may include the ORCID identifers as follows.  F. Smith, https://orcid.org/0000-0001-2345-6789; B. Jones, https://orcid.org/0000-0009-8765-4321}

\backsection[Author contributions]{Authors may include details of the contributions made by each author to the manuscript'}

\appendix

\section{Experimental setup}\label{appA}

As illustrated in Fig. \ref{fig1}(a), a vertical tank was designed to study bubble deformation and breakup in HIT. This vertical tank consists of two main components: an octagonal test section and an upward-facing jet array system to generate HIT. Fig. \ref{fig1}(a) inset shows the jet array which is designed similarly as the one by \cite{masuk2019v}. The jet array features 21 circular nozzles with a separation distance of 50.4 mm (as indicated by the black arrow). The nozzle diameter is 8 mm, and these nozzles are turned on and off randomly at a frequency of 0.5 Hz to eliminate any large-scale flows. In this work, water jets were fired at 7 m/s, and 10 out of 21 jets on average were kept on at a time to maximize the turbulence intensity. The jet array only introduces momentum not mass into the test section as the same amount of water injected is also taken back through those 16 squared through holes.

The octagonal test section is 1 m tall and 23 cm in diameter as an inscribed circle. The wall is made of 25.4 mm thick acrylic sheets for optical access. Four high-speed cameras working at 1280$\times$800 resolution and 5000 frames per second were used to image the view volume (red shaded area in Fig. \ref{fig1}(a)), and four designated LED panels provided diffused light to cast shadows of bubbles and tracer particles onto the camera’s imaging plane as shown in Fig. \ref{fig1}(b). The bubbles were injected directly into the test section via a needle with an inner diameter of 5 mm, and the air was supplied by a stainless steel gas line (represented by the thick gray line in Fig. \ref{fig1}(a)). The tip of the needle is located at around 6 times the nozzle-to-nozzle separation distance above the jay array, where jets are fully mixed and the generated turbulence becomes homogeneous and isotropic as suggested by \cite{tan2023scalings}. By carefully adjusting the gas flow rate using a syringe pump, a single bubble with a diameter ranging from 7 to 13 mm was generated in each run in order to avoid large bubble clusters generated from breakups blocking camera views.

\section{Bubble interfacial velocity and curvature}\label{appB}
The octagonal test section as well as the multiple-camera arrangement enabled us to perform 3D reconstruction of the bubble geometry, from which the bubble trajectory can be then obtained simultaneously. The reconstruction of bubble geometry is performed by adopting the visual hull method \citep{masuk2019robust}. In this process, the bubble 3D geometry is reconstructed by calculating the intersection of the cone-like volume extruded from the bubble silhouettes extracted from each camera. Based on the the time-resolved bubble geometry, the interfacial velocity $\boldsymbol{u}_f$ can be determined following four steps: (i) the bubble geometry is first smoothed by applying a filter with a filter length of around $0.2D$. Applying such a filter inevitably removes the high wave number structures on the bubble interface. However, structures below this scale could not be discerned from reconstruction uncertainty anyway; (ii) in order to focus on the interfacial dynamics, the translational motion of the bubble is subtracted based on the bubble trajectories so that the center of the bubble at different times coincides with one another; (iii) the displacement of the bubble interface on each vertex can be obtained by tracking vertices between two consecutive frames using the nearest neighbor algorithm. By dividing this displacement by the time delay between the two frames, the velocity on each vertex can be calculated; (iv) this velocity is then projected to the normal direction of the interface to acquire the interfacial velocity $\boldsymbol{u}_f$ as the tangential component of the interfacial velocity is not measurable. In addition to $\boldsymbol{u}_f$, the mean curvature $\kappa=(\kappa_1+\kappa_2)/2$ of the bubble interface was also calculated along each bubble trajectory based on the method proposed by \cite{rusinkiewicz2004estimating}. Here, $\kappa_1$ and $\kappa_2$ are the maximum and minimum principal curvature of the interface, respectively. Fig. \ref{fig1}(c) shows examples of the reconstructed bubble geometry, the distribution of $u_f$, and the distribution of $\kappa$ of the same bubble.

\section{Structure function}\label{app_struct_func}

\begin{figure}
    \centering
    \includegraphics[width=0.5\linewidth]{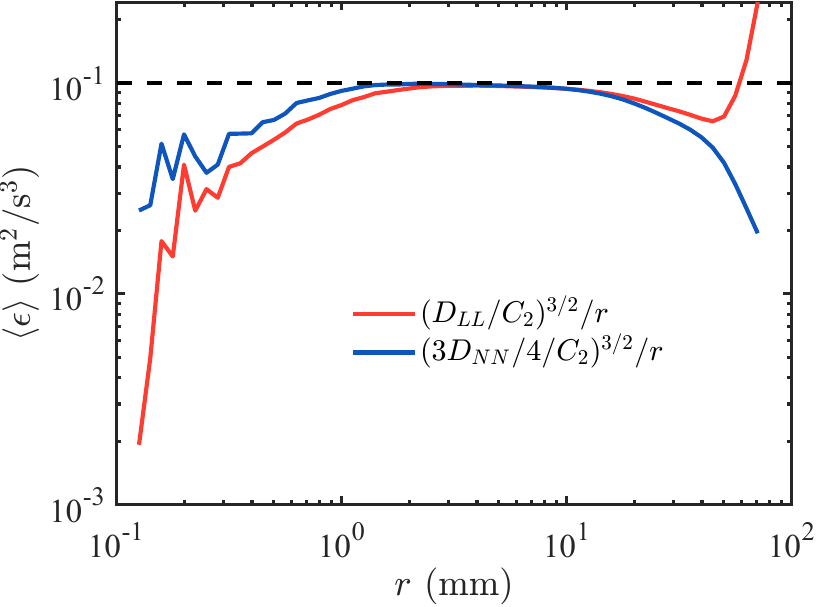}
    \caption{Estimated energy dissipation rate $\langle\epsilon\rangle$ based on the longitudinal (red) and transverse (blue) second-order structure function. The dashed line represents $\langle\epsilon\rangle=0.1$ m$^2/$s$^3$.}
    \label{fig_struct_func}
\end{figure}

For the continuous phase, by seeding 60 $\mu$m tracer particles (with the Stokes number $St=0.06$) into the flow, 3D trajectories of tracers were reconstructed by performing Lagrangian particle tracking using the in-house OpenLPT code \citep{tan2020}. These trajectories were further smoothed by convoluting them with Gaussian kernels \citep{mordant2004experimental,ni2012lagrangian} from which the tracer velocity was obtained along the trajectory. To better quantify the flow properties, we calculated the second-order structure function for both the longitudinal $D_{LL}(r)$ and transverse $D_{NN}(r)$ components, based on the processed tracer velocity following a similar procedure as in \cite{masuk2021simultaneous} , where $r$ is the separation distance between a pair of tracer particles. It is noted that, based on the Kolmogorov theory \citep{kolmogorov1941local}, the second-order structure function in the inertial range can be estimated using the energy dissipation rate $\langle\epsilon\rangle$, i.e., $D_{LL}=C_2(\langle\epsilon\rangle r)^{2/3}$ and $D_{NN}=(4/3)C_2(\langle\epsilon\rangle r)^{2/3}$, where $C_2\approx 2$ is the Kolmogorov constant. These relations provide us a way to estimate $\langle\epsilon\rangle$ following $\langle\epsilon\rangle=(D_{LL}/C_2)^{3/2}/r$ and $\langle\epsilon\rangle=(3D_{NN}/4/C_2)^{3/2}/r$, as shown in Fig. \ref{fig_struct_func}. It is evident that both $D_{LL}$ and $D_{NN}$ seem to collapse well in the inertial range where the plateau is shown. Within the inertial range, $\langle\epsilon\rangle$ can be estimated by extracting the magnitude of the plateau, i.e., $\langle\epsilon\rangle\approx 0.1$ m$^2/$s$^3$. Note that $\langle\epsilon\rangle$ estimated based on both $D_{LL}$ and $D_{NN}$ are consistent with each other, suggesting the flow is close to homogeneous and isotropic.

Based on $\langle\epsilon\rangle$ obtained from Fig. \ref{fig_struct_func}, the Kolmogorov length scale $\eta$ and Kolmogorov time scale $\tau_\eta$ can be estimated by following $\eta=(\nu^3/\langle\epsilon\rangle)^{1/4}=56$ $\mu$m and $\tau_\eta=(\nu/\langle\epsilon\rangle)^{1/2}=3.2$ ms, where $\nu$ is the kinematic viscosity of the flow. Given the range of bubble sizes $D$ generated, the Weber number following $We=\rho C_2 (\langle\epsilon\rangle D)^{2/3} D /\sigma$ ranges from $1.5$ to $4.3$, indicating turbulence is sufficiently strong to drive the deformation and breakup of bubbles in our system. Here, $\rho$ is the density of the continuous phase, and $\sigma$ is the surface tension coefficient.

\bibliographystyle{jfm}
\bibliography{ref}



\end{document}